\documentclass[vecphys]{svmult}

\usepackage{graphicx}        % standard LaTeX graphics tool
\usepackage[bottom]{footmisc}% places footnotes at page bottom

\begin{document}

\title*{The Bipolar Engines of post-AGB stars: Transient Dynamos and Common Envelopes}
\author{J. Nordhaus\inst{1}\inst{2}\and E. G. Blackman\inst{1}\inst{2}}
\institute{Dept. of Physics and Astronomy, University of Rochester, Rochester NY 14627
\and Laboratory for Laser Energetics, University of Rochester, Rochester NY 14623 \texttt{nordhaus@pas.rochester.edu}}
\maketitle
\begin{abstract}
The physical mechanism(s) responsible for transitioning from a spherical Asymptotic Giant Branch (AGB) star to an asymmetric post-AGB (pAGB) object is poorly understood.  In particular, excess momenta in the outflows of pAGB objects suggest that a binary may be required to supply an additional source of energy and angular momentum.  The extraction of rotational energy from the engine is likely fundamental and may be facilitated if a dynamo is operating in the interior.  In this regard, single star magnetic outflow models have been proposed as mechanisms for producing and shaping PNe, however these models neglect the back-reaction of the large-scale magnetic field on the flow.  

Here we present a transient $\alpha-\Omega$ dynamo operating in the envelope of an AGB star in (1) an isolated setting and (2) a common envelope in which the secondary is a low-mass companion in-spiraling in the AGB interior.  The back reaction of the fields on the shear is included and differential rotation and rotation deplete via turbulent dissipation and Poynting flux.  For an isolated star, the shear must be resupplied in order to sufficiently sustain the dynamo.  We comment on the energy requirements that convection must satisfy to accomplish this.  For the common envelope case, a robust dynamo can result as the companion provides an additional source of energy and angular momentum.

\end{abstract}

\section{Binaries and Additional Momenta}
Post-AGB objects, and correspondingly, many PNe, exhibit extreme asymmetry in the form of collimated jets and/or bipolar structures.  However the physical processes responsible for shaping pAGB outflows have remained elusive for almost two decades.  Magnetic field detection and maser emission in many objects have sustained interest in magnetic launching and collimation mechanisms \cite{Bains,Jordan,Sabin}.  In particular, a magnetically collimated jet in an evolved star further suggests a dynamical role of the magnetic field in pAGB evolution \cite{Vlemmings}.  

Single star magnetic outflow models have been proposed as mechanisms for shaping and producing pPNe/PNe \cite{Pascoli,Blackman,Garcia-Segura}.  However, these models neglect the catastrophic quenching which occurs when the large-scale field back-reacts on the flow.  As the magnetic field grows, differential rotation is drained rapidly ($\leq 100$ yrs) and results in the termination of the dynamo, making it difficult for isolated stars to produce observed asymmetries.  On the other hand, if convection can resupply differential rotation in an AGB star, then an envelope dynamo in an isolated star may be viable.  

If single star models fail, the observed bipolarity may instead be the result of energy and angular momentum supplied by a binary companion.  This is supported by recent radial velocity surveys suggesting that many, if not all, PNe harbor binaries or incurred a binary interaction \cite{De Marco,Sorensen}.  Additionally, most pAGB systems exhibit extreme momentum excesses ($\sim 10^2-10^4$ times larger than that supplied by radiation pressure) \cite{Bujarrabal}.  A binary companion may provide a natural source of additional momentum, especially if the interaction results in a common envelope (CE) phase \cite{JTN2006}.

Here, we review results of ongoing theoretical studies which aim to understand how low-mass companions (planets, brown-dwarfs and low-mass main sequence stars) produce asymmetries in evolved stars \cite{JTN2006,JTN2007}.  We focus on common envelope evolution in which a low-mass companion is engulfed  during the AGB phase.  Three distinct ejection scenarios are identified, leading to qualitatively different mass outflow consequences (see \S \ref{sec2}).  In particular, we investigate a magnetic model in which the companion spins-up the common envelope, driving a dynamo in the interior.  We compare these results to a dynamo operating in an isolated AGB star.  Constraints are placed upon the isolated star scenario for it to be a viable engine in producing bipolarity (see \S \ref{sec3}).  A common envelope dynamo, on the other hand, is robust and can drive asymmetries for a range of outflow types and companion masses (see \S \ref{sec4}).  We conclude and comment in \S \ref{sec5}.

\section{Common Envelopes: The case of low-mass companions}
\label{sec2}
Roche lobe overflow in close binary systems can result in both stars immersed in a common envelope \cite{Paczynski,IbenLivio}.  Once inside, a drag force generated by velocity differences between the primary envelope and companion, induces in-spiral.  If the mass ratio of the system is low ($<$ $0.1$), the secondary in-spirals towards the primary core.  

\begin{figure}%use [h!] for 'here'
\centering
\resizebox{6.2cm}{!}{\includegraphics{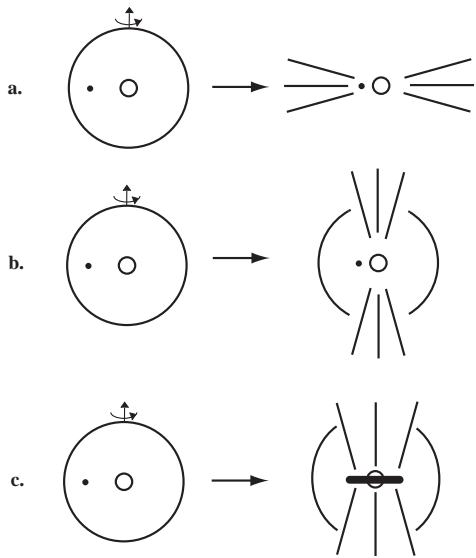}}
\caption{Three possible outcomes of our CE evolution with a low-mass secondary.  (a) The companion becomes embedded in the stellar envelope directly ejecting the envelope equatorially. (b) The envelope is spun-up causing it to differentially rotate.  An envelope dynamo ensues, unbinding material along the poles. (c) The companion is shredded into an accretion disc around the core.  The disc drives an outflow which unbinds the envelope poloidally \cite{JTN2006}.}
\label{Fig1}
\end{figure}

During evolution off the main-sequence, expansion of the primary's envelope may engulf a companion either by tidal capture or directly during the expansion.  In-spiral of a low-mass companion (planet, brown dwarf, low-mass MS star) into an AGB star envelope was investigated in \cite{JTN2006}.  The transfer of energy and angular momentum can eject the AGB envelope and influence outflow direction.   Three mass ejection scenarios are presented in Fig. \ref{Fig1}.  In scenario (a), the companion injects enough orbital energy and angular momentum into the AGB envelope to directly unbind it and eject material in an equatorial torus.  In scenario (b), the companion spins-up the envelope causing it to differentially rotate and drive a dynamo in the interior.  The dynamo can unbind the envelope and eject material poloidally.  In scenario (c), the companion is shredded into an accretion disc around the core.  The disc drives an outflow, unbinding the envelope and launching material along the poles.  For more on these scenarios, see \cite{Blackmanproceedings} (these proceedings).  In this paper, we focus on (b) in the case of an isolated AGB star and one in which a low-mass companion has spun-up the envelope during a common envelope phase.  We refer the reader to \cite{JTN2007} for in-depth details.

\section{Isolated AGB Dynamo}
\label{sec3}
An unresolved issue in magnetic PNe progenitor models is whether an isolated AGB star can sustain the necessary field strengths and corresponding Poynting flux to unbind the envelope and produce collimated outflows.  We present results of a dynamical model in which the back-reaction of field growth on the flow is incorporated for a $3$ $M_\odot$ AGB star.  Angular momentum is conserved on spherical shells as the star evolves off the main sequence and provides the initial differential rotation profile.  As the large-scale magnetic field amplifies, shear energy is correspondingly drained.  However, in an isolated AGB star, there is too little shear energy to generate strong magnetic fields and the dynamo terminates after $\sim20$ yrs \cite{JTN2007}.  A constant differential rotation profile must be established, in order to sustain the dynamo until the aggregate Poynting flux is dynamically important \cite{JTN2007}.  This circumstance occurs in the sun as convection re-seeds differential rotation through the $\lambda$-effect \cite{Rudiger}.

\begin{figure}
\centering
\includegraphics[height=8.0cm]{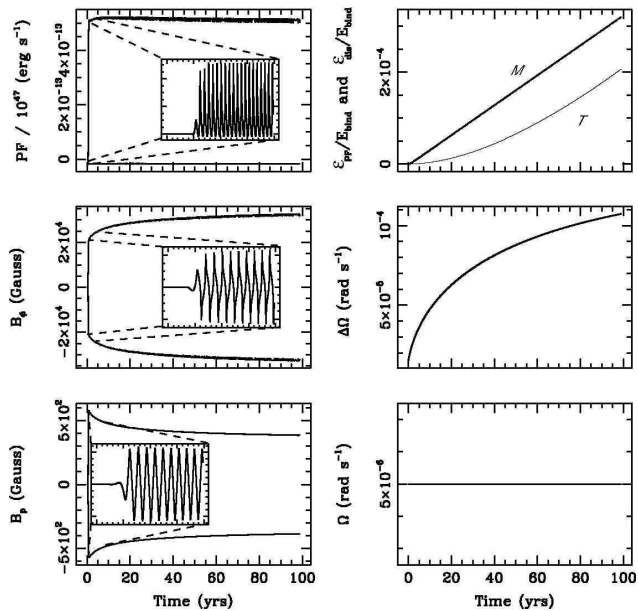}
\caption{Convective resupply results in a steady-state differential rotation profile.  Left column: poloidal field, $B_p$ (bottom), toroidal field, $B_\phi$ (middle) and Poynting flux (top).  Insets represent the time evolution to 2 yrs.  Right column: rotation, $\Omega$ (bottom), differential rotation, $\Delta\Omega$ (middle) and the fraction of the envelope binding energy supplied via heat and Poynting flux (top).}
\label{resupply}       
\end{figure}

Although it remains to be established dynamically if a similar effect occurs in evolved stars, by analogy to the solar case, we allow a fraction of the turbulent energy cascade to resupply shear \cite{JTN2007}.  Additionally, we keep the rotation at the interface between the convective and shear zones fixed.  This is physically equivalent to storing the Poynting flux in the interface region (i.e. magnetic buoyancy is negligible).  If the field is trapped, Poynting flux does not emerge from the layer and thus, does not spin-down the envelope.  A single star dynamo may be viable when the following two conditions are satisfied: (i.) convection resupplies differential rotation (ii.) Poynting flux is stored inside the envelope.  

For our $3$ $M_\odot$ AGB star, under the above two conditions, a steady-state dynamo is established when $\sim1\%$ of the turbulent cascade energy reinforces the shear (see Fig. \ref{resupply}) .  The peak Poynting flux is sustained at $\sim5\times10^{34}$ erg s$^{-1}$ and accumulates until the stored field supplies enough energy to unbind the envelope at the end of the AGB phase ($\sim 10^5$ yrs).  If the above two conditions are met, then an isolated star dynamo may be viable as a mechanism for producing bipolar outflows in pPNe.

\section{Common Envelope Dynamo}
\label{sec4}
Unlike single star models, a binary companion via a common envelope phase offers an additional source of energy and angular momentum which is otherwise unavailable \cite{JTN2006}.  A common envelope is advantageous as the in-spiral time is fast ($<1$ yr) and thus, energy and angular momentum are delivered quickly.  We consider low-mass (planets, brown dwarfs, low-mass main sequence stars) embedded in the envelope of our model AGB star, and use the gravitational potential energy released by the secondary during in-spiral to spin up spherical shells.  The  dynamical equations and model are presented in detail in \cite{JTN2007}.  

The in-spiral of even low-mass companions can significantly spin-up the envelope.  Higher mass secondaries supply enough orbital energy to spin-up the envelope above its Keplerian value at a given radius.  The rotational energy will then redistribute via outward mass transfer until Keplerian rotation is re-established.  Ideally, the differential rotation profile should be solved for self-consistenly with this effect becoming particularly important as soon as the rotation rate exceeds the sound speed.  We have not incorporated this redistribution explicitly and hence, our approach of redistributing the excess rotational energy in the inner regions is approximate.  Nevertheless, the required energy and angular momentum are present in the CE phase.  

Our solutions are categorized by the relative amount of energy supply to the envelope by the time-integrated heat and Poynting flux.  A key parameter in determining whether the induced outflow would be thermally or magnetically driven is the turbulent magnetic Prandtl number, $Pr_p\equiv \beta_\phi/\beta_p <1$.  In our model AGB star, the convective zone is highly turbulent while the shear layer is weakly so.  Therefore, we parameterize the turbulent diffusion coefficients in each region such that $\beta_\phi\gg\beta_p$.  The turbulent diffusion coefficient in the differential rotation zone governs how far the poloidal component of the field can diffuse into the shear zone in a cycle period.  The further into the shear zone the poloidal field can penetrate, the greater the shear energy that can be extracted and utilized via the dynamo.  In contrast, $\beta_p$ governs how much heat is generated by turbulent dissipation.  Thus, it is the interplay of these two quantities, coupled with the companion mass that ultimately determines whether a model is thermally or magnetically driven.

\begin{figure}[h!]
\centering
\includegraphics[height=8cm]{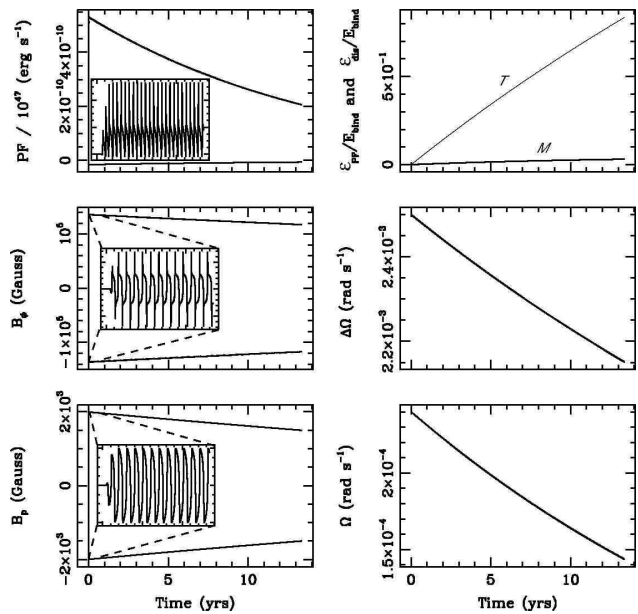}
\caption{Interface dynamo resulting from the in-spiral of a $0.02$ $M_\odot$ brown dwarf.  For this model, heat from turbulent dissipation is the dominant source of energy and is the primary driver in unbinding the envelope.  Such an outflow is expected to be quasi-spherical.}
\label{thermal}
\end{figure}

We present an example of a thermally driven model in Fig. \ref{thermal}.  In this case, a $0.02$ $M_\odot$ brown dwarf in-spirals through the AGB envelope.  Even though the dynamo is operating and the magnetic field is amplified, in this case, heat supplies the required energy to unbind the envelope.  For $Pr_p = 10^{-4}$ (Fig. \ref{thermal}), the decay of the shear energy and toroidal field is long ($\sim 25$ yrs) an occurs over several thousand cycle periods.  If heat is the primary driver in mediating the transition from progenitor to pAGB, the resulting outflow is probably quasi-spherical and may not be the mechanism responsible for the production of bipolar PNe.

\begin{figure}[h!]
\centering
\includegraphics[height=8cm]{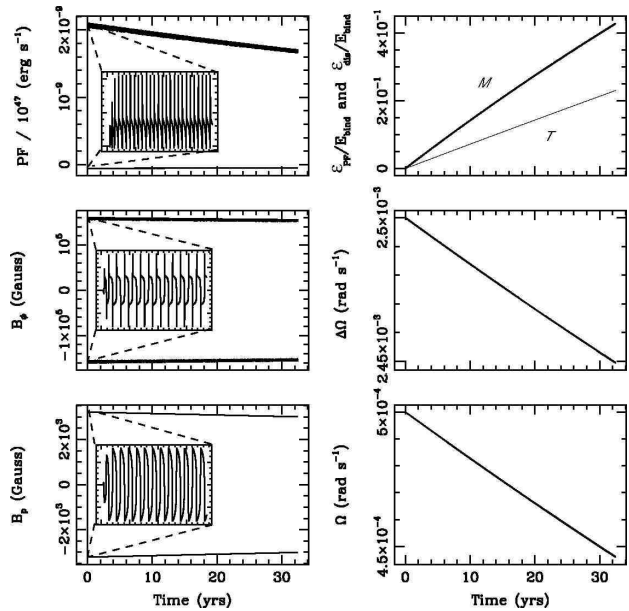}
\caption{For this model, the secondary is a $0.05$ $M_\odot$ brown dwarf.  Strong differential rotation in the interior generates a Poynting flux which unbinds the envelope after $\sim 100$ yrs.  The resulting outflow is expected to be bipolar and collimated.}
\label{magnetic}
\end{figure}

In contrast to a thermally driven model, we also identify a situation in which the time-integrated Poynting flux is large enough to unbind the envelope.  In Fig. \ref{magnetic}, a $0.05$ $M_\odot$ brown dwarf in-spirals through our AGB star and spins-up the envelope.  For this model, $Pr_p=10^{-6}$ with the corresponding Poynting flux decaying in $\sim 100$ yrs.  The peak field strengths are comparable to those obtained in Fig. \ref{thermal}, however the lower $Pr_p$ results in less differential rotation energy being converted into heat.  Instead, the dynamo lifetime is longer and the aggregate Poynting flux larger.  We therefore, identify this as a magnetically driven model.  In this situation, the outflow is expected to be magnetically launched, collimated and bipolar.  In addition, for some models, the launch may be explosive and could be responsible for the production of ansae where the burst-times are generally $\sim 100-300$ yrs.  It may also be possible to produce steady, magnetically collimated winds from these results for a range of time-scales.  To fully investigate this problem, future research should link magnetic field amplification to the physics of the jet-launch in a self-consistent manner.   

Both magnetically and thermally driven models can be produced for a range of companion masses and diffusion coefficients.  The resultant outflows for the two cases are quite different.  For dynamo driven winds, the launching and shaping of the outflow may occur close to the core.  Such an outflow is expected to be collimated, predominately poloidal and may be responsible for shaping features in Abell 63 \cite{Mitchell}.  On the other hand, if heat is the primary transitioning mechanism between the AGB and pAGB phase, the resulting outflow is probably quasi-spherical and may be responsible for producing elliptical or spherical pPNe/PNe.  Bipolar, magnetically collimated pPNe/PNe could be the result of common envelope, magnetically driven models.

\section{Conclusions}
\label{sec5}
Extraction of rotational energy is likely fundamental to the formation of multipolar PPNe and PNe.  Magnetic dynamos can play an intermediary role in facilitating the extraction of rotational energy, however, previous models have neglected the crucial effect of the back-reaction of field growth on the shear.  We have presented results of dynamos incorporating this effect for both isolated stars and common envelope systems.  For an isolated star, stringent conditions (including resupply of shear and storage of Poynting flux) must be met if an isolated dynamo is to be viable in producing bipolar PNe.  

Common envelope evolution is robust in supplying the requisite energy and angular momentum needed to produce strong, bipolar magnetic fields in the pAGB phase.  We have discussed two paradigms: magnetically driven (bipolar, collimated) and thermally driven (quasi-spherical) which may be responsible for shaping during the pAGB phase.

%%%%%%%%%%%%%%%%%%%%%%%%%%%%%%%%%%%%%%%%%%%%%%%%%%%%%%%%%%%%%%%%%%%%%%  }

%%%%%%%%%%%%%%%%%%%%%%%%%%%%%%%%%%%%%%%%%%%%%%%%%%%%%%%%%%%%%%%%%%%%%%

\end{document}